\documentclass{cimento}

\usepackage{amsfonts,amsmath,amssymb,bm,graphicx,cite,makeidx,multicol,color}

\title{New neutrino spin oscillations in moving matter and magnetic fields}

\author{ A.~I.
Studenikin\from{ins:MSU}\from{ins:JINR}\thanks{studenik@srd.sinp.msu.ru}}

\instlist{
\inst{ins:MSU} Department of Theoretical Physics, Lomonosov Moscow State
University, 119991 Moscow,  Russia \inst{ins:JINR} Joint Institute
for Nuclear Research, 141980 Dubna, Russia }

\PACSes{\PACSit{13.15.+g} {Neutrinos interactions}
        \PACSit{95.30.Cq} {Elementary particles in astrophysics} }
\shorttitle{New effects in neutrino oscillations}
\begin{document}

\maketitle

\begin{abstract}
After a short introduction to the history of neutrino mixing and oscillations, we celebrate the 50th anniversary of the first analytical expression for the neutrino evolution probability obtained by Gribov and Pontecorvo in 1969 by the  discussion of two new phenomena in neutrino oscillations: 1) the emergence of the neutrino spin and spin-flavour oscillations engendered by the transversal in respect to neutrino propagation matter currents, and 2) an inherent interplay of neutrino oscillations on the vacuum $\omega_{vac}=\frac{\Delta m^2}{4p}$ and magnetic $\omega_{B}=\mu B_{\perp}$ frequencies in the case of neutrino propagation in a constance magnetics field. We also predict a new phenomena of the modification of the flavour neutrino oscillations probability in moving matter that can be engendered by a non-vanishing   matter transversal current 
${\bf {j}}_{\perp}=n {\bf {v}}_{\perp}$.
\end{abstract}
\section{Introduction}
\label{Sec_2}

We give a fast look on several new aspects in the phenomena
of neutrino oscillations in moving matter and a magnetic field discussed
in our papers \cite{Pustoshny:2018jxb,Studenikin:2004bu} and \cite{Popov:2019nkr,Popov:2018seq}. By the present short notes
we celebrate the 50th anniversary of the first analytic expression for the neutrino
oscillation probability obtained in the paper \cite{Gribov:1968kq}
by Gribov and Pontecorvo in 1969. Then we also predict a new phenomena of the modification of the flavour neutrino oscillations probability in moving matter that can be engendered by a non-vanishing   matter transversal current $\bm{j}_{\perp}=n \bm {v}_{\perp}$.

\section{A brief history of neutrino mixing and oscillations}

\subsection{Neutrino flavour oscillations in vacuum and matter}

 The story of the neutrino mixing and oscillations started with two papers by Bruno Pontecorvo \cite{Pontecorvo:1957cp,Pontecorvo:1957qd} where the above mentioned effects have been discussed for the first time. A history of neutrino mixing and oscillations can be found in \cite{Bilenky:2016pep}.
 In \cite{Pontecorvo:1957cp} Pontecorvo has indicated that if the neutrino charge were not conserved then the transition between a neutrino and antineutrino would become possible in vacuum. In \cite{Pontecorvo:1957qd} Pontecorvo has even directly introduced a phenomenon of neutrino mixing.
In 1962, just after the discovery of the second flavour neutrino, the effect of neutrino mixing was discussed in \cite{Maki:1962mu} where
the fields of the weak neutrinos $\nu_e$ and $\nu_{\mu}$ were connected with the neutrinos mass states
$\nu_1$ and $\nu_2$ by the unitary mixing matrix $U$ that can be parametrized by the mixing angle $\theta$ and
\begin{equation}
\nu_e = \nu_1 \cos \theta + \nu_2 \sin \theta, \ \ \  \nu_\mu = - \nu_1 \sin \theta + \nu_2 \cos \theta .
\end{equation}

The theory of neutrino mixing and oscillations was further developed in \cite{Gribov:1968kq, Bilenky:1975tb} with actual calculations of neutrino beam evolution. In \cite{Wolfenstein:1977ue} the effect of neutrino interaction with matter of a constant density on neutrino flavour mixing and oscillations was investigated. The existence of resonant amplification of neutrino mixing  (the MSW effect) when a neutrino flux propagates through a medium with varying density was predicted in \cite{Mikheev:1986gs}. Generalization of the MSW resonance condition
for the case of moving matter was derived in \cite{Grigoriev:2002zr}.

The tedious studies, both experimental and theoretical, over the last 60 years has been honored by the Nobel Prize of 2015 awarded to Arthur McDonald and Takaaki Kajita for the discovery of neutrino oscillations, which shows that neutrinos have mass.

{\it{Neutrino spin oscillations in magnetic fields.}}
The straightforward consequences of neutrino nonzero mass is the prediction \cite{Fujikawa:1980yx} that neutrinos can have nonzero magnetic moments. Studies of neutrino magnetic moments and the related phenomena attract a reasonable interest in literature.  The values of neutrino magnetic moments are constrained in the terrestrial laboratory experiments and in the astrophysical considerations (see, for instance,
\cite{Beda:2012zz,Borexino:2017fbd} and \cite{Raffelt:1990pj}).

Massive neutrinos participate in electromagnetic interactions. The recent review on this topic is given in
  \cite{Giunti:2014ixa} (the upgrate can be found in \cite{Studenikin:2018vnp}). One of the most important phenomenon of nontrivial neutrino electromagnetic
interactions is the neutrino magnetic moment precession and the corresponding
spin oscillations in presence of external
electromagnetic fields. The later effect has been studied in numerous papers published
during several passed decades.

Within this scope the neutrino spin oscillations $\nu^{L}\Leftrightarrow \nu^{R}$ induced by the neutrino magnetic moment interaction with the transversal magnetic field ${\bf B}_{\perp}$ was first considered in \cite{Cisneros:1970nq}. Then spin-flavor oscillations $\nu^{L}_{e}\Leftrightarrow \nu^{R}_{\mu}$ in ${\bf B}_{\perp}$ in vacuum were discussed
in \cite{Schechter:1981hw}, the importance of the matter effect was emphasized in \cite{Okun:1986hi}.
The effect of the resonant amplification of neutrino spin oscillations in ${\bf B}_{\perp}$ in the presence of matter was proposed in \cite{Akhmedov:1988uk,Lim:1987tk}. Neutrino spin oscillations in a magnetic field with
account for the effect of moving matter was studied in \cite{Lobanov:2001ar}. A possibility to establish
conditions for the resonance in neutrino spin oscillations by the effect of matter motion discussed in
\cite{Studenikin:2004bu}. The impact of the longitudinal magnetic field ${\bf B}_{||}$ was discussed in \cite{Akhmedov:1988hd}. The neutrino spin oscillations in the presence of constant twisting magnetic field
were considered in \cite{Vidal:1990fr, Smirnov:1991ia, Akhmedov:1993sh,
Likhachev:1990ki,Dvornikov:2007aj,Dmitriev:2015ega}.

In \cite{Egorov:1999ah} neutrino spin oscillations were considered in the presence of  arbitrary constant electromagnetic fields $F_{\mu \nu}$. Neutrino spin oscillations in the presence of the field of circular and linearly polarized electromagnetic waves and superposition of an electromagnetic wave and constant magnetic field were considered in \cite{Lobanov:2001ar, Dvornikov:2001ez}.
The effect of the parametric resonance in neutrino oscillations in periodically varying electromagnetic fields
was studied in \cite{Dvornikov:2004en}.

More general case of neutrino spin evolution in the case when neutrino is subjected to general types of non-derative interactions with external scalar $s$, pseoudoscalar $\pi$, vector $V_{\mu}$, axial-vector $A_{\mu}$, tensor $T_{\mu\nu}$ and pseudotensor $\Pi_{\mu\nu}$ fields was considered in
\cite{Dvornikov:2002rs}. From the  general neutrino spin evolution equation, obtained in \cite{Dvornikov:2002rs}, it follows that neither scalar $s$ nor pseudoscalar $\pi$ nor vector $V_{\mu}$ fields can induce neutrino spin evolution. On the contrary, within the general consideration of neutrino spin evolution it was shown that electromagnetic (tensor) and weak (axial-vector) interactions can contribute to the neutrino spin evolution.

We have also considered in details \cite{Fabbricatore:2016nec,Studenikin:2016zdx,Kurashvili:2017zab} neutrino mixing and oscillations in arbitrary constant magnetic field that have  ${\bf B}_{\perp}$ and ${\bf B}_{||}$ nonzero components and derived an explicit expressions for the effective neutrino magnetic moments
for the flavour neutrinos in terms of the corresponding magnetic moments
introduced in the neutrino mass basis.

\section{New effects in neutrino flavour, spin and spin-flavour oscillations
in a constant magnetic field}\label{osc_B}

Recently a new approach  to description of neutrino spin and spin-flavor oscillations in the presence of an arbitrary constant magnetic field has been  developed
(see also \cite{Dmitriev:2015ega,Studenikin:2016zdx}). Within the new approach exact quantum stationary states  are used for classification of neutrino spin states, rather than the neutrino helicity states that have been  used for this purpose within the customary approach in many published papers. Recall that the helicity states are not stationary in the presence of a magnetic field. It has been shown \cite{Popov:2019nkr,Popov:2018seq},
in particular, that in the presence of the transversal magnetic field $B_{\perp}$ for a given choice of parameters (the  energy and magnetic moments of neutrinos and strength of the magnetic field)  the amplitude of the
flavour oscillations $\nu_e^L \Leftrightarrow \nu_{\mu}^L$ at the vacuum frequency
$\omega_{vac}=\frac{\Delta m^2}{4p}$ is modulated
by the magnetic field frequency $\omega_{B}=\mu B_{\perp}$:
\begin{equation}\label{fl_simp}
P^{(B_{\perp})}_{\nu_{e}^L \rightarrow \nu_{\mu}^L}(t) = \left( 1 - \sin^2(\mu B_{\perp}t) \right) \sin^2 2\theta \sin^2 \frac{\Delta m^2}{4p}t   = \left(1 - P_{\nu_{e}^L \rightarrow \nu_{e}^R}^{cust}\right)
P_{\nu_{e}^L \rightarrow \nu_{\mu}^L}^{cust},
\end{equation}
here $\mu$ is the effective magnetic moment of the electron neutrino and
it is supposed that the following relations between diagonal and transition magnetic
moments in the neutrino mass basis are valid: $\mu_1 = \mu_2, \ \ \mu_{ij}=0, \ i\neq j$.
The customary expression
\begin{equation}\label{flavour_cust}
 P_{\nu_{e}^L \rightarrow \nu_{\mu}^L}^{cust}(t)=  \sin^2 2\theta \sin^2 \frac{\Delta m^2}{4p}t
 \end{equation}
for the neutrino flavour oscillation probability in vacuum in the presence of the transversal
field $B_{\perp}$ is modified by the factor $1 - P_{\nu_{e}^L \rightarrow \nu_{e}^R}^{cust}$. Since the transition magnetic moment in the flavour basis is absent in the case $\mu_1 = \mu_2$, the process $\nu_e^L \rightarrow \nu_e^R$ is the only way for spin flip, and then $1 - P_{\nu_{e}^L \rightarrow \nu_{e}^R}^{cust}$ should be interpreted as the probability of not changing the neutrino spin polarization. And consequently, this multiplier subtracts the contribution of neutrinos $\nu^{R}_e$ with the opposite polarization
providing the survival of the only contribution from the direct neutrino flavour oscillations $\nu_e^L \Leftrightarrow \nu_{\mu}^L$.

Similar results on the important influence of the transversal magnetic field on amplitudes of various types of neutrino oscillations were obtained earlier \cite{Kurashvili:2017zab} on the basis of the exact solution of the effective equation for neutrino evolution in the presence of a magnetic field, which accounts for four neutrino species corresponding to two different flavor states with positive and negative helicities.

Consider the probability of the neutrino spin-flavour oscillations $\nu_e^L \leftrightarrow \nu_{\mu}^R$.
In the case $\mu_1 = \mu_2 = \mu$ we have \cite{Popov:2019nkr,Popov:2018seq}:
\begin{equation}\label{spin_flavour_simplified}
P_{\nu_e^L \rightarrow \nu_{\mu}^R}(t) =  \sin^2(\mu B_{\perp} t)\sin^2 2\theta \sin^2\frac{\Delta m^2}{4p}t.
\end{equation}
The obtained expression (\ref {spin_flavour_simplified}) for the probability can be expressed as a product of two probabilities derived within the customary  two-neutrino-states approach
\begin{equation}\label{PP}
P_{\nu_e^L \rightarrow \nu_{\mu}^R} (t)=
P_{\nu_{e}^L \rightarrow \nu_{\mu}^L}^{cust} (t) P_{\nu_{e}^L \rightarrow \nu_{e}^R}^{cust}(t),
\end{equation}
where the probability $P_{\nu_e^L \rightarrow \nu_{\mu}^L}^{cust}(t)$ given by
(\ref {flavour_cust}) and
 the usual expression for the neutrino spin oscillation probability
\begin{equation}\label{P_mu_B}
P_{\nu_{e}^L \rightarrow \nu_{e}^R}^{cust}(t) = \sin^2(\mu B_{\perp}t),
\end{equation}
are just the probabilities obtained in the customary approach. A
similar  neutrino spin-flavour oscillations (for the Majorana case) as a
two-step neutrino conversion processes were considered in \cite{Akhmedov:2002mf}. Since the probability of neutrino spin-flavour oscillations was supposed to be small, this effect was calculated \cite{Akhmedov:2002mf} within perturbation theory.

Finally, in the case of the spin oscillations $\nu_{e}^L \rightarrow \nu_{e}^R$
in the transversal magnetic field $B_{\perp}$ within the the developed approach the
effect of neutrino mixing is accounted for that leads to the modification of the
customary expression $P_{\nu_{e}^L \rightarrow \nu_{e}^R}^{cust}(t)$ for the probability by the factor
$1 - P_{\nu_{e}^{L} \rightarrow \nu_{\mu}^{L}}^{cust}$:
\begin{eqnarray}
\label{spin_simplified}
P_{\nu_{e}^L \rightarrow \nu_{e}^R} = \left[ 1 - \sin^2 2\theta \sin^2\left(\frac{\Delta m^2}{4p}t\right) \right]\sin^2(\mu B_{\perp}t).
\end{eqnarray}

The interplay between different oscillations, that follows from the obtained expressions (\ref{fl_simp}), (\ref{spin_flavour_simplified}) and (\ref{spin_simplified}) for the
oscillation probabilities,  gives rise to interesting phenomena (see also \cite{Popov:2019nkr,Popov:2018seq}):

1) the amplitude modulation of the probability of flavour oscillations $\nu_e^L \rightarrow \nu_{\mu}^L$ in the transversal magnetic field with the magnetic frequency $\omega_{B}=\mu B_{\perp}$ (in the case $\mu_1 = \mu_2$) and more complicated dependence  on the harmonic functions with $\omega_{B}$ for $\mu_1 \neq \mu_2$;

2) the dependence of the spin oscillation probability $P_{\nu_{e}^L \rightarrow \nu_{e}^R}$ on the mass square difference $\Delta m^2$;

3) the appearance of the spin-flavour oscillations in the case $\mu_1=\mu_2$ and $\mu_{12}=0$, the transition goes through the two-step processes $\nu_e^L \rightarrow \nu_{\mu}^L \rightarrow \nu_{\mu}^R$ and $\nu_e^L \rightarrow \nu_{e}^R \rightarrow \nu_{\mu}^R$.

As a result, we predict modifications of the neutrino oscillation patterns that might provide new
 important phenomenological consequences in case of neutrinos propagation in extreme
 astrophysical environments where magnetic fields are present.
\section{New effect of neutrino spin and spin-flavour oscillations in transversal matter current}
\label{osc_j}

Following the discussion in \cite{Studenikin:2004bu} consider, as an example,  an electron neutrino spin precession in the case when neutrinos with the Standard Model interaction are propagating through moving and polarized matter composed of electrons (electron gas) in the presence of an electromagnetic field given by the electromagnetic-field tensor $F_{\mu \nu}=({\bf E}, {\bf B})$.
To derive the neutrino spin oscillation probability in the transversal matter current we use the generalized Bargmann-Michel-Telegdi equation that describes  the evolution of the
three-di\-men\-sio\-nal neutrino spin vector $\bf S $,
\begin{equation}\label{S}  {d{\bf S}
\over dt}={2 \mu \over \gamma} \Big[ {\bm S} \times ({\bm
B}_0+{\bm M}_0) \Big],
\end{equation}
where the magnetic field $\bf{B}_0$ in the neutrino rest frame.
The matter term ${\bf M}_0$ in Eq. (\ref{S}) is composed of the transversal ${\bf  M}{_{0_{\parallel}}}$
and longitudinal  ${\bf  M}_{0_{\perp}}$ parts in respect to the direction of the neutrino propagation ,
\begin{equation}
{\bf M}_0=\bf {M}{_{0_{\parallel}}}+{\bf M}_{0_{\perp}},
\label{M_0}
\end{equation}
where
\begin{equation}\label{M_0_parallel}
{\bf M}_{0_{\parallel}}={{{G}_{F}(1+4\sin^2 \theta _W) \over {2\sqrt{2}\mu \sqrt {1- v_{e}^{2}}
 }}}n_{0}\left(1-{{\bf v}_e
{\bm\beta} \over {1- {\gamma^{-2}}}} \right)\gamma{\bm\beta},
\end{equation}
\begin{equation}\label{M_0_perp}
 {\bf M}_{0_{\perp}}=-{{{G}_{F}(1+4\sin^2 \theta _W) \over {2\sqrt{2}\mu \sqrt {1-v_{e}^{2}}}}}
 n_{0}{\bf v}_{e_{\perp}},
\end{equation}
here $n_0=n_{e}\sqrt {1-v^{2}_{e}}$ is the invariant number density of the background particles in
the rest reference frame of matter, $\gamma = (1-\beta^2)^{-{1 \over 2}}$,
$\bm{\beta}$ is the neutrino velocity and ${\bf v}_e$ the velocity of matter.

The probability of the neutrino spin oscillations engendered by the transversal matter
current (by the term ${\bf M}_{0_{\perp}}$) was obtained for the first time in \cite{Studenikin:2004bu}
\begin{equation}\label{ver2}
P_{\nu^{L}_{e} \rightarrow \nu^{R}_{e}} (x)=\sin^{2} 2\theta_\textmd{eff}
\sin^{2}{\pi x \over L_\textmd{eff}},  \ \ \sin^{2} 2\theta_\textmd{eff}={M_{0\perp}^2\over
{M_{0\parallel}^2 +M_{0\perp}^2}}, \ \ \
L_\textmd{eff}={\pi \over {\mu M_0}}\gamma.
\end{equation}
Note that the possibility of neutrino spin oscillations due to interaction with matter of a rotating astrophysical object was also indicated in \cite{Studenikin:2004tv}.

The introduced above semiclassical theory of neutrino spin oscillations engendered by
the transversal matter currents has been recently appended by the direct quantum treatment
of the phenomenon \cite{Pustoshny:2018jxb,Studenikin:2016iwq, Studenikin:2017pag,Studenikin:2017mdh}
\

Consider two flavour neutrinos with two possible helicities
$\nu_{f}= (\nu_{e}^{+}, \nu_{e}^{-}, \nu_{\mu}^{+}, \nu_{\mu}^{-})^T$ in moving matter
composed of neutrons. Each of the flavour neutrinos is a superposition of the neutrino mass states,
\begin{equation}\label{transformations}
    \begin{array}{c}
  \nu_{e}^{\pm} =\nu_{1}^{\pm}\cos\theta+\nu_{2}^{\pm}\sin\theta,\ \ \ \ \
  \nu_{\mu}^{\pm}=-\nu_{1}^{\pm}\sin\theta+\nu_{2}^{\pm}\cos\theta.
\end{array}
\end{equation}
The corresponding neutrino evolution equation is
\begin{equation}\label{schred_eq_fl}
  i\dfrac{d}{dt}\nu_{f}=H^{f}_{v}\nu_{f},
\end{equation}
where the evolution Hamiltonian reads
\begin{equation}\label{H_v}
H^f_v=n\tilde{G}\left( \begin{matrix}
0& (\frac{\eta}{\gamma})_{ee}v_{\perp} &0 &(\frac{\eta}{\gamma})_{e\mu}v_{\perp}\\
(\frac{\eta}{\gamma})_{ee}v_{\perp} & 2(1-v_{\parallel}) &(\frac{\eta}{\gamma})_{e\mu}v_{\perp}&
0\\
0& (\frac{\eta}{\gamma})_{e\mu}v_{\perp} &0 &(\frac{\eta}{\gamma})_{\mu \mu}v_{\perp}\\
(\frac{\eta}{\gamma})_{e\mu}v_{\perp} & 0 &(\frac{\eta}{\gamma})_{\mu\mu}v_{\perp}& 2(1-v_{\parallel})\\
\end{matrix} \right),
\end{equation}
and
\begin{equation}\label{eta}
\Big(\frac{\eta}{\gamma}\Big)_{ee}=
\frac{\cos^2\theta}{\gamma_{11}}+\frac{\sin^2\theta}{\gamma_{22}}, \ \
 \Big(\frac{\eta}{\gamma}\Big)_{\mu\mu}=
\frac{\sin^2\theta}{\gamma_{11}}+\frac{\cos^2\theta}{\gamma_{22}}, \ \
\Big(\frac{\eta}{\gamma}\Big)_{e\mu}=
\frac{\sin 2\theta}{\tilde{\gamma}_{21}},
 \end{equation}
where ${\widetilde{\gamma}_{2 1}}^{-1}=\frac{1}{2}\big(
\gamma_{2}^{-1}-\gamma_{1}^{-1}\big),\ \ \gamma_{\alpha}^{-1}=\frac{m_\alpha}{p^{\nu}_0}$,
$p^{\nu}_0$ is neutrino energy and $\alpha = 1,2 $.

Consider the initial neutrino state
$\nu_{e}^L$ moving in the background with the magnetic field
${\bm B} = {\bm B}_{\parallel}+{\bm B}_{\perp}$ and
nonzero matter current ${\bm j} = {\bm j}_{\parallel}+{\bm j}_{\perp}$.
One of the possible modes of neutrino transitions with the change of helicity is
$\nu_{e}^L\Leftarrow (j_{\perp}, B_{\perp}) \Rightarrow \nu_{e}^R$. Here we restrict our
consideration to the to the The corresponding
oscillations are governed by the evolution equation \cite{Pustoshny:2018jxb}
\begin{eqnarray}\label{nu_L_nu_R}
	i\frac{d}{dt} \begin{pmatrix}\nu^L_{e} \\ \nu^R_{e} \\  \end{pmatrix}=
	 \left( \begin{matrix}
	(\frac{\mu}{\gamma})_{ee}{B_{||}}+{\widetilde{G}}n(1-{\bm v}{\bm \beta})\\
	 \mu_{ee}B_{\perp}+ (\frac{\eta}{\gamma})_{ee}{\widetilde{G}}nv_{\perp}
	 \end{matrix} \right.
	 \left. \begin{matrix} \mu_{ee}B_{\perp} + (\frac{\eta}{\gamma})_{ee}{\widetilde{G}}nv_{\perp}  \\
	  - (\frac{\mu}{\gamma})_{ee}{B_{||}} -{\widetilde{G}}n(1-{\bm v}{\bm \beta} \end{matrix} \right)
	\begin{pmatrix}\nu^L_{e} \\ \nu^R_{e} \\ \end{pmatrix}.
\end{eqnarray}
Here we constraint our consideration to the binary neutrino transitions and corresponding
oscillations between pairs of the neutrino states.
The oscillation $\nu^L_{e} \Leftarrow (j_{\perp}, B_{\perp}) \Rightarrow \nu^R_{e}$ probability
is given by
\begin{equation}\label{prob_oscillations}
P^{(j_{\perp},B_{\perp})}_{\nu^L_{e} \rightarrow \nu^R_{e}} (x)={E^2_\textmd{eff} \over
{E^{2}_\textmd{eff}+\Delta^{2}_\textmd{eff}}}
\sin^{2}{\pi x \over L_\textmd{eff}}, \ \ \ L_\textmd{eff}={\pi \over \sqrt{E^{2}_\textmd{eff}+\Delta^{2}_\textmd{eff}}},
\end{equation}
where
\begin{equation}\label{E}
E_{eff}= \Big|\mu_{ee}\bm{B}_{\perp} + \Big(\frac{\eta}{\gamma}\Big)
_{ee}{\widetilde{G}}n\bm{v}_{\perp}  \Big|, \ \ \
\Delta_{eff}= \Big|\Big(\frac{\mu}{\gamma}\Big)_{ee}{\bm{B}_{||}}+
{\widetilde{G}}n(1-{\bm v}{\bm \beta}){\bm {\beta}} \Big|.
\end{equation}
From (\ref{prob_oscillations}) and (\ref{E}) it is clearly seen that even in
the absence of the transversal magnetic field
the neutrino spin oscillations $\nu_{e}^L\Leftarrow (j_{\perp}) \Rightarrow \nu_{e}^R$
can be generated by the neutrino interaction with the transversal matter current
$\bm{j}_{\perp}=n\bm{v}_{\perp}$.

In a quite analogous way the transversal matter current can engender the neutrino
spin-flavour conversion and the corresponding oscillations $\nu_{e}^L\Leftarrow (j_{\perp}) \Rightarrow\nu_{\mu}^R$ between two different flavour
states with opposite spin orientations. Note that for apperance of this effect
there is no need for a magnetic field (obviously, the effect is neutrino
magnetic moments independent). For the neutrino evolution $\nu_{e}^L\Leftarrow (j_{\perp}) \Rightarrow\nu_{\mu}^R$ in the case $B=0$
from (\ref{H_v}) we get
\begin{eqnarray}
	i\frac{d}{dt} \begin{pmatrix}\nu^L_{e} \\ \nu^R_{\mu} \\  \end{pmatrix}=
	\left( \begin{matrix}
	 -\Delta M+{\widetilde{G}}n(1-{\bm v}{\bm \beta}) \\
	  (\frac{\eta}{\gamma})_{e\mu}{\widetilde{G}}nv_{\perp} \end{matrix} \right.
	 \left. \begin{matrix}  (\frac{\eta}{\gamma})_{e\mu}{\widetilde{G}}nv_{\perp}  \\
	 \Delta M -{\widetilde{G}}n(1-{\bm v}{\bm \beta})
		\end{matrix} \right)
	\begin{pmatrix}\nu^L_{e} \\ \nu^R_{\mu} \\ \end{pmatrix},
\end{eqnarray}
where $\Delta M=\frac{\Delta m ^2 \cos 2\theta}{4 p^{\nu}_0 }$.
The probability $P_{\nu_{e}^L \rightarrow \nu_{\mu}^R}^{(j_{\perp})}$ of the neutrino spin-flavour
oscillations $\nu^L_{e} \Leftarrow (j_{\perp}) \Rightarrow \nu^R_{\mu}$ is given by the same equation (\ref{prob_oscillations}), but now
\begin{eqnarray}\label{E_1_delta_1}
E_{eff}= \Big|\Big(\frac{\eta}{\gamma}\Big)_{e\mu}{\widetilde{G}}n{v}_{\perp}  \Big|, \ \
\Delta_{eff}
=\Big|\Delta M-{\widetilde{G}}n(1-{\bm v}{\bm \beta})  \Big|.
\end{eqnarray}
From (\ref{E_1_delta_1}) it follows that
the neutrino spin-flavour oscillations $\nu^L_{e} \Leftarrow (j_{\perp}) \Rightarrow \nu^R_{\mu}$
can be generated by the neutrino interaction with the transversal matter current
$\bm{j}_{\perp}=n\bm{v}_{\perp}$.

\section{New effect: Modification of
  flavour neutrino oscillations probability in moving matter by
  transversal matter current}

  From the previous discussions it follows that in the presence of a magnetic field (see Section
  \ref{osc_B} and also \cite{Popov:2019nkr,Popov:2018seq} for details) it is not possible to consider the neutrino flavour and spin oscillations
  as separate phenomena. On the contrary, there is an inherent communication between two.
  In particular, the amplitude of the neutrino flavour oscillations is modulated by the
  magnetic frequency $\omega_{B}=\mu B_{\perp}$. The main result of Section \ref{osc_j}
  (see also \cite{Studenikin:2004bu,Pustoshny:2018jxb})
  is the conclusion on the equal role that the transversal magnetic field ${\bm B}_{\perp}$ and the
  transversal matter current $\bm{j}_{\perp}$ plays in generation of the neutrino spin and spin-flavour oscillations.
  From these observations \textbf{ we predict a new phenomena of the modification of
  the flavour neutrino oscillations probability in moving matter that can be engendered by a non-vanishing
  matter transversal current $\bm{j}_{\perp}=n \bm {v}_{\perp}$ }.

  The flavour neutrino oscillation probability
 accounting for this effect can be expressed, under quite reasonable conditions,  as follows:
  \begin{equation}\label{flav_mod_current}
P^{(j_{||}+j_{\perp})}_{\nu_{e}^L \rightarrow \nu_{\mu}^L} (t) =  \left(1 - P_{\nu_{e}^L \rightarrow \nu_{e}^R}^{(j_{\perp})} - P_{\nu_{e}^L \rightarrow \nu_{\mu}^R}^{(j_{\perp})}\right)
P_{\nu_{e}^L \rightarrow \nu_{\mu}^L}^{(j_{||})},
\end{equation}
where
\begin{equation}
P_{\nu_{e}^L \rightarrow \nu_{\mu}^L}^{(j_{||})}(t) =
\sin^2 2\theta _{eff}\sin^2 \omega_{eff}t, \ \ \omega_{eff}=\frac{\Delta m^2 _{eff}}{4p_{0}^\nu},
\end{equation}
is the flavour oscillation probability in moving matter \cite{Grigoriev:2002zr}, $\theta _{eff}$
and $\Delta m^2 _{eff}$  are the corresponding quantities modified by the presence of moving matter
(note that in the definition of $\theta _{eff}$
and $\Delta m^2 _{eff}$ only the longitudinal component of matter motion matters).
For the probability of the neutrino spin oscillations  engendered by the transversal current $\bm {j}_{\perp}$
from (\ref{nu_L_nu_R}) we get
\begin{equation}
P^{j_{\perp}}_{\nu_{e}^L \rightarrow \nu_{e}^R}(t)=
\frac{\Big(\frac{\eta}{\gamma}\Big)_{ee}^2 {v}^{2}_{\perp}}
{\Big(\frac{\eta}{\gamma}\Big)_{ee}^2 {v}^{2}_{\perp}+
(1-{\bm v}{\bm \beta})^2}
\sin^2\omega^{j_{\perp}}_{ee}t.
\end{equation}
For the corresponding probability of the neutrino spin-flavour  oscillations due to $\bm {j}_{\perp}$
from (\ref{E_1_delta_1}) we get
\begin{equation}
P^{j_{\perp}}_{\nu_{e}^L \rightarrow \nu_{\mu}^R}(t)=
\frac{\Big(\frac{\eta}{\gamma}\Big)_{e\mu}^2 {v}^{2}_{\perp}}
{\Big(\frac{\eta}{\gamma}\Big)_{e\mu}^2 {v}^{2}_{\perp}+
\Big(\frac{\Delta M}{{\widetilde{G}}n} - (1-{\bm v}
{\bm \beta})\Big)^2}
\sin^2\omega^{j_{\perp}}_{e\mu}t.
\end{equation}
The discussed new effect of the modification of the
flavour oscillations $\nu_{e}^L\Leftarrow (j_{||}, j_{\perp})\Rightarrow \nu_{\mu}^L$ probability
is the result of an interplay of oscillations on a customary
flavour oscillation frequency in moving matter $\omega_{eff}$ and
two additional oscillations with changing the neutrino polarization,
the neutrino spin $\nu_{e}^L\Leftarrow (j_{\perp}) \Rightarrow \nu_{e}^R$ and spin-flavour
$\nu_{e}^L\Leftarrow (j_{\perp}) \Rightarrow \nu_{\mu}^R$ oscilations, that are governed by two
characteristic frequencies
\begin{equation}\label{omega_1}
  \omega^{j_{\perp}}_{ee}={\widetilde{G}}n{\sqrt{\Big(\frac{\eta}{\gamma}\Big)_{ee}^2 {v}^{2}_{\perp} +
  (1-{\bm v}{\bm \beta})^2}}
\end{equation}
and
\begin{equation}\label{omega_2}
  \omega^{j_{\perp}}_{e\mu}={\widetilde{G}}n
  {\sqrt{\Big(\frac{\eta}{\gamma}\Big)_{e\mu}^2 {v}^{2}_{\perp}
  +
  \Big(\frac{\Delta M}{{\widetilde{G}}n}-(1-{\bm v}{\bm \beta})\Big)^2}}.
\end{equation}
The discussed interplay of oscillations on different frequencies can lead to important consequences in the case when neutrino fluxes are propagating in astrophysical environments peculiar to rotating compact objects or dense jets of matter.

{\it Acknowledgements.} The authors thanks Giorgio
Bellettini, Giorgio Chiarelli, Mario Greco and Gino Isidori for the
kind invitation to participate in the XXXIII Recontres de Physique de
La Vallee D'Aoste on Results and Perspectives in Particle Physics and
also thanks all the organizers for their hospitality in La Thuile.

\end{document}